# Nonrelativistic Green's function for systems with position-dependent mass

A. D. Alhaidari

*Physics Department, King Fahd University of Petroleum & Minerals, Box 5047, Dhahran 31261, Saudi Arabia*
E-mail: haidari@mailaps.org

Given a spatially dependent mass we obtain the two-point Green's function for exactly solvable nonrelativistic problems. This is accomplished by mapping the wave equation for these systems into well-known exactly solvable Schrödinger equations with constant mass using point canonical transformation. The one-dimensional oscillator class is considered and examples are given for several mass distributions.

## I. INTRODUCTION

Quantum systems with spatially dependent effective mass were found to be very useful models for investigating the physical properties of various microstructures and semiconductor interfaces in condensed matter. Special applications of these models are carried out in the study of electronic properties of semiconductors,[1] quantum wells and quantum dots,[2] 3He clusters,[3] quantum liquids,[4] graded alloys and semiconductor heterostructures,[5] …etc. These applications stimulated a lot of work in the literature on the development of methods and techniques for studying systems with mass that depends on position. Recently, several contributions have emerged that give solutions of the wave equation for such systems. The one dimensional Schrödinger equation with smooth mass and potential steps was solved exactly in Ref. 6. The formalism of supersymmetric quantum mechanics was extended to include position dependent mass.[7] Shape invariance was also addressed in this setting and the energy spectra were obtained for several examples. A class of solutions was obtained explicitly for such systems with equi-spaced spectra.[8] Coordinate transformation in supersymmetric quantum mechanics were used in Ref. 9 to generate isospectral potentials for systems with position dependent mass. The ordering ambiguity of the mass and momentum operators and its effect on the exact solutions was addressed in Ref. 10 where several examples are considered. so(2,1) Lie algebra as a spectrum generating algebra and as a potential algebra was used in Ref. 11 to obtain exact solutions of the effective mass wave equation. Point canonical transformation (PCT) was used in Ref. 12 to obtain the energy spectra and wavefunctions explicitly for a large class of problems in one and three dimensions. A class of quasi-exactly solvable problems with effective mass was presented in Ref. 13 where the wavefunctions are obtained in terms of orthogonal polynomials satisfying mass dependent recurrence relation.

In all cited work above the main concern is in obtaining the energy spectra and/or wavefunctions for these systems once the position dependent mass function is given. Moreover, exact solvability requirements result in constraints on the potential functions for the given mass distribution. On the other hand, the Green's functions for such systems, which are of prime significance in the calculation of physical processes, did not receive adequate attention. We are aware of only one contribution that dealt with the Green's functions for systems with position dependent mass: In 1995, Chetouani, Dekar, and Hammann used path integral formulation to relate the constant mass Green's function to that of position dependent mass.[14] This was done on a formal level with explicit results in the two cases of step and rectangular-barrier potential and mass functions. In this article, we extend the PCT method used in Ref. 12 for obtaining the energy spectra and wavefunctions of such systems to the calculation of the two-point Green's function. The basic idea behind the PCT method is as follows.[15] Starting with a problem whose solution (exact, quasi-exact, or conditionally-exact) is known, then applying to it coordinate transformation that preserves the canonical form of the wave equation will map it into other solvable problems. The canonical constraint on the coordinate transformation generates classes of these solvable problems.

In Sec. II, we start with the one-dimensional time-independent Schrödinger equation satisfied by the Green's function for a system with constant mass (the reference problem). Applying to it PCT maps it into the wave equation for the Green's function of a system with position dependent mass. The canonical constraint defines the coordinate transformation in terms of the given mass function. It gives, as well, the potential functions for solvable systems with this position dependent mass that belong to the class of the reference (constant mass) problem. A correspondence among the physical parameters of the two problems will also be generated. In Sec. III, the formalism is implemented on the one-dimensional oscillator class and examples are given for several mass distributions.

## II. ACTION OF THE PCT MAP ON THE GREEN'S FUNCTION

The momentum operator no longer commutes with the mass since the latter depends on position. In the majority of work done on the subject the following symmetric ordering of mass and momentum, in the kinetic energy part of the Hamiltonian, is adopted almost unanimously:

$$H = \frac{1}{2}\left[\vec{P}\frac{1}{M(\vec{r})}\vec{P}\right] + V(\vec{r}) = -\frac{\hbar^2}{2m_0}\left[\vec{\nabla}\frac{1}{m(\vec{r})}\vec{\nabla}\right] + V(\vec{r}) \tag{2.1}$$

where $m(\vec{r})$ and $V(\vec{r})$ are real functions of the configuration space coordinates. Using atomic units ($m_0 = \hbar = 1$), this will result in the following time independent wave equation in one dimension

$$\left\{\frac{d^2}{dx^2} - \frac{m'}{m}\frac{d}{dx} - 2m[V(x) - E]\right\}\phi(x) = 0 \tag{2.2}$$

where $E$ is the energy eigenvalue and $m' \equiv dm/dx$. The Green's function (resolvent operator) $g_E$ associated with this problem is formally defined as $(H-E)^{-1}$, where $E$ does not belong to the discrete spectrum of the Hamiltonian $H$. It satisfies the following inhomogeneous equation:

$$\left\{\frac{d^2}{dx^2} - \frac{m'}{m}\frac{d}{dx} - 2m[V(x) - E]\right\} g_E(x,\bar{x}) = 2m\delta(x-\bar{x}) \tag{2.3}$$

On the other hand, the one-dimensional equation satisfied by the two-point Green's function for a system with constant mass, potential function $\mathcal{V}$, and energy $\mathcal{E}$ reads

$$\left\{\frac{d^2}{dy^2} - 2[\mathcal{V}(y) - \mathcal{E}]\right\} \mathcal{G}_{\mathcal{E}}(y, \overline{y}) = 2\delta(y - \overline{y}) \tag{2.4}$$

We apply to this last equation the following transformation:
$$y = q(x), \quad \mathcal{G}_{\mathcal{E}}(y, \overline{y}) = p(x) g_E(x, \overline{x}) p^*(\overline{x}) \tag{2.5}$$

If the result is a mapping into Eq. (2.3) then this transformation will be referred to as "point canonical transformation (PCT)". Now the action of (2.5), for real functions, on Eq. (2.4) maps it into the following equation

$$\left\{\frac{d^2}{dx^2} + \left(2\frac{p'}{p} - \frac{q''}{q'}\right)\frac{d}{dx} + \left(\frac{p''}{p} - \frac{q''}{q'}\frac{p'}{p}\right) - 2(q')^2[\mathcal{V}(q(x)) - \mathcal{E}]\right\} g_E(x, \overline{x}) = \frac{2(q')^2}{p(x) p(\overline{x})} \delta(q(x) - q(\overline{x}))$$

By identifying this with Eq. (2.3) and using the relation $q' \delta(q(x) - q(\overline{x})) = \delta(x - \overline{x})$ we obtain the following conditions on the transformation (2.5) to be a PCT:

$$p(x) = \sqrt{q'/m} \tag{2.6}$$

$$V(x) - E = \frac{(q')^2}{m}[\mathcal{V}(q) - \mathcal{E}] + \frac{1}{4m}[F(m) - F(q')] \tag{2.7}$$

where $F(z) = z''/z - \frac{3}{2}(z'/z)^2$. Given a position dependent mass $m(x)$, Eq. (2.7) is a constraint relating the potential function $V(x)$ to the transformation function $q(x)$ for a given class defined by the reference potential $\mathcal{V}(y)$. Therefore, for each choice of potential $V(x)$ there will be an associated PCT function $q(x)$ satisfying Eq. (2.7). Once $q(x)$ is determined then so is $p(x)$ as it is given by Eq. (2.6). Consequently, the Green's function $g_E(x, \overline{x})$ for the position-dependent mass system will be given by (2.5) in terms of the known reference Green's function $\mathcal{G}_{\mathcal{E}}(y, \overline{y})$ as

$$g_E(x, \overline{x}) = \sqrt{m(x) m(\overline{x}) / q'(x) q'(\overline{x})} \; \mathcal{G}_{\mathcal{E}}(q(x), q(\overline{x})) \tag{2.8}$$

Moreover, a correspondence map will also be generated by Eq. (2.7) relating the physical parameters of the reference problem (e.g., $\mathcal{E}$) to those of the variable mass problem (e.g., $E$).

Our strategy for solving the constraint equation (2.7) is by choosing a PCT function $q(x)$ that will result in a position independent term on the right side of Eq. (2.7) which will be identified with the constant energy term $E$ on the left. To this end we consider the following two possibilities:

(a) The first is $(q')^2 = m$ giving the PCT function $q(x) = \tau \mu(x)$, where $\tau$ is a length scale positive parameter and

$$\mu(x) = (1/\tau) \int \sqrt{m(x)} \, dx \tag{2.9}$$

For a given mass distribution $m(x)$, this choice of PCT function, when substituted in Eq. (2.7), results in the following energy and potential function:

$$\begin{aligned} E &= \mathcal{E} \\ V(x) &= \mathcal{V}(\tau \mu(x)) + \frac{1}{8 m(x)} G(m(x)) \end{aligned} \tag{2.10}$$

where $G(z) = z''/z - \frac{7}{4}(z'/z)^2$. It will also give the following sought after two-point Green's function:

$$g_E(x, \overline{x}) = [m(x) m(\overline{x})]^{1/4} \mathcal{G}_{\mathcal{E}}(\tau \mu(x), \tau \mu(\overline{x})) \tag{2.11}$$

(b) The second PCT function is obtained by taking $(q')^2 \mathcal{V}(q) = \pm m/\sigma^2$, where $\sigma$ is another real parameter. This choice gives $q(x) = R^{-1}(\tau \mu(x)/\sigma)$, where $R(y) = \int \sqrt{\pm \mathcal{V}(y)} \, dy$, and results in the following:

$$E = \mp 1/\sigma^2$$

$$V(x) = \mp \frac{\mathcal{E}/\sigma^2}{\mathcal{V}(q(x))} + \frac{1}{8m(x)} G(m(x))$$

$$\pm \frac{1}{8\sigma^2 \mathcal{V}(q(x))} \left\{ \frac{\mathcal{V}''(q(x))}{\mathcal{V}(q(x))} - \frac{5}{4} \left[ \frac{\mathcal{V}'(q(x))}{\mathcal{V}(q(x))} \right]^2 \right\} \quad (2.12)$$

$$g_E(x, \bar{x}) = \sigma \left[ m(x) m(\bar{x}) \mathcal{V}(q(x)) \mathcal{V}(q(\bar{x})) \right]^{1/4} \mathcal{G}_\varepsilon(q(x), q(\bar{x}))$$

where $\mathcal{V}' \equiv d\mathcal{V}(q)/dq$. Requiring that the first term in the potential expression above be independent of energy (through $\sigma$) will result in a constraint that relates the parameter $\sigma$ to the reference energy $\mathcal{E}$. However, the last term in $V(x)$ will always be independent of $\sigma$. This is due to the fact that this term comes from $F(q')$ in the general relation (2.7) which is homogeneous in $q$ with zero degree.

It is to be noted, however, that other choices of $q(x)$ could also be found that might produce a constant term on the right hand side of Eq. (2.7), thus resulting in other solutions. However, we are contented here with the two solutions obtained above. In the following section we use this development to obtain the nonrelativistic two-point Green's function for several systems with different position-dependent mass that belong to the oscillator class.

## III. OSCILLATOR CLASS GREEN'S FUNCTIONS

In this section we apply the PCT method development above to obtain the non-relativistic two-point Green's function for several systems with position-dependent mass in the oscillator class where $\mathcal{V}(y) = \frac{1}{2} \omega^4 y^2$ and $\omega$ being the oscillator frequency. In this case, the PCT choice $(q')^2 = m$ gives $q(x) = \omega^{-1} \mu(x)$, where $\mu(x)$ is the dimensionless integral in (2.9) with the length scale parameter $\tau$ taken equals to $1/\omega$. The potential function obtained using Eq. (2.10) is

$$V(x) = \tfrac{1}{2} \omega^2 \mu(x)^2 + \tfrac{1}{8m(x)} G(m(x)) \quad (3.1)$$

On the other hand, the PCT choice $(q')^2 \mathcal{V}(q) = m/\sigma^2$ gives $q(x)^2 = (2\sqrt{2}/\sigma \omega^3) \mu(x)$ and results in the following potential function by using Eq. (2.12)

$$V(x) = -\frac{\mathcal{E}/\sqrt{2}\sigma}{\omega \mu(x)} + \frac{1}{8m(x)} G(m(x)) - \frac{3}{32} \frac{\omega^2}{\mu(x)^2}$$

To eliminate the energy dependence in the first term of this potential we require that $\sigma$ be linearly proportional to $\mathcal{E}$. From dimensional arguments and using the available parameters in the problem, we redefine $\sigma$ as $\sigma = (\sqrt{2}/\lambda \omega^3) \mathcal{E}$, where $\lambda$ is a dimensionless real parameter. Consequently, for this PCT choice, which now reads $q(x)^2 = (2\lambda/\mathcal{E})\mu(x)$, Eq. (2.12) gives the following energy, potential, and Green's function:

$$E = -\tfrac{1}{2}\lambda^2\omega^6/\mathcal{E}^2$$
$$V(x) = -\tfrac{\lambda}{2}\omega^2\mu(x)^{-1} - \tfrac{3}{32}\omega^2\mu(x)^{-2} + \tfrac{1}{8m(x)}G(m(x)) \qquad (3.2)$$
$$g_E(x,\bar{x}) = \sqrt{2\omega}\,(-2E)^{-1/4}\left[m(x)m(\bar{x})\mu(x)\mu(\bar{x})\right]^{1/4}\mathcal{G}_\mathcal{E}(q(x),q(\bar{x}))$$

Now to proceed beyond this point we need to compute the reference Green's function $\mathcal{G}_\mathcal{E}(y,\bar{y})$ for the constant mass one-dimensional oscillator. This Green's function is well known. One may consult Ref. 16 for a recent review. In one of its representations, we could write it as

$$\mathcal{G}_\mathcal{E}(y,\bar{y}) = \tfrac{1}{W_\mathcal{E}}\psi_\mathcal{E}^-(y_<)\psi_\mathcal{E}^+(y_>) \qquad (3.3)$$

where $y_>$ ($y_<$) is the larger (smaller) of $y$ and $\bar{y}$. $\psi_\mathcal{E}^\pm(y)$ are two independent solutions of the Schrödinger wave equation $\{d^2/dy^2 - 2[\mathcal{V}(y) - \mathcal{E}]\}\psi_\mathcal{E}^\pm(y) = 0$ which are regular at the boundary limits of $y_>$ ($y_<$), respectively. The Wronskian of these two solutions is written as

$$W_\mathcal{E} = \psi_\mathcal{E}^+(y)\frac{d\psi_\mathcal{E}^-(y)}{dy} - \psi_\mathcal{E}^-(y)\frac{d\psi_\mathcal{E}^+(y)}{dy} \qquad (3.4)$$

which is independent of $y$ as can be verified by differentiating with respect to $y$ and using the wave equation. The explicit form of $\mathcal{G}_\mathcal{E}(y,\bar{y})$ depends on whether the one-dimensional configuration space is taken to be the whole real line $y \in (-\infty,+\infty)$ or only half the line $y \in [0,+\infty)$.[16] For the whole line it reads

$$\overset{\leftrightarrow}{\mathcal{G}}_\mathcal{E}(y,\bar{y}) = \tfrac{2}{\sqrt{\pi}\omega^2}\Gamma\!\left(\tfrac{3}{4} - \tfrac{\mathcal{E}}{2\omega^2}\right)\tfrac{1}{\sqrt{y\bar{y}}} \times$$
$$\left[\mathcal{M}_{\mathcal{E}/2\omega^2,1/4}(\omega^2 y_<^2) + \tfrac{1/4}{\sqrt{\pi}}\Gamma\!\left(\tfrac{1}{4} - \tfrac{\mathcal{E}}{2\omega^2}\right)\mathcal{W}_{\mathcal{E}/2\omega^2,1/4}(\omega^2 y_<^2)\right]\mathcal{W}_{\mathcal{E}/2\omega^2,1/4}(\omega^2 y_>^2) \qquad (3.5)$$

However, in the case of the semi-infinite real line ($y \geq 0$) we write it as follows:

$$\overset{\rightarrow}{\mathcal{G}}_\mathcal{E}(y,\bar{y}) = \tfrac{2}{\sqrt{\pi}\omega^2}\Gamma\!\left(\tfrac{3}{4} - \tfrac{\mathcal{E}}{2\omega^2}\right)\tfrac{1}{\sqrt{y\bar{y}}}\mathcal{M}_{\mathcal{E}/2\omega^2,1/4}(\omega^2 y_<^2)\mathcal{W}_{\mathcal{E}/2\omega^2,1/4}(\omega^2 y_>^2) \qquad (3.6)$$

where $\Gamma$ is the gamma function, $\mathcal{M}_{a,b}$ and $\mathcal{W}_{a,b}$ are the Whittaker functions of the first and second kind, respectively.[17] They are defined in terms of the confluent hyper-geometric functions as

$$\mathcal{M}_{a,b}(z) = z^{b+1/2}e^{-z/2}{}_1F_1(b-a+\tfrac{1}{2};2b+1;z)$$
$$\mathcal{W}_{a,b}(z) = \tfrac{\Gamma(-2b)}{\Gamma(1/2-b-a)}\mathcal{M}_{a,b}(z) + \tfrac{\Gamma(2b)}{\Gamma(1/2+b-a)}\mathcal{M}_{a,-b}(z) \qquad (3.7)$$

Substituting the above reference Green's functions (3.5) and (3.6) into either one of the two formulas for $g_E(x,\bar{x})$ in Eq. (2.11) or Eq. (3.2) gives the sought after Green's functions. That is, we end up with the following two possibilities:

(a) $q(x) = \omega^{-1}\mu(x)$:

$$V(x) = \tfrac{1}{2}\omega^2 \mu(x)^2 + U(x)$$

$$\Leftrightarrow \bar{g}_E(x,\bar{x}) = \frac{2}{\sqrt{\pi}}\Gamma\left(\tfrac{3}{4} - \tfrac{E}{2\omega^2}\right)\frac{[m(x)m(\bar{x})]^{1/4}}{\omega\sqrt{\mu(x)\mu(\bar{x})}} \times$$

$$\left[\mathcal{M}_{E/2\omega^2, 1/4}(\mu(x_<)^2) + \tfrac{1/4}{\sqrt{\pi}}\Gamma\left(\tfrac{1}{4} - \tfrac{E}{2\omega^2}\right)\mathcal{W}_{E/2\omega^2, 1/4}(\mu(x_<)^2)\right]\mathcal{W}_{E/2\omega^2, 1/4}(\mu(x_>)^2)$$

$$\Rightarrow \bar{g}_E(x,\bar{x}) = \frac{2}{\sqrt{\pi}}\Gamma\left(\tfrac{3}{4} - \tfrac{E}{2\omega^2}\right)\frac{[m(x)m(\bar{x})]^{1/4}}{\omega\sqrt{\mu(x)\mu(\bar{x})}}\mathcal{M}_{E/2\omega^2, 1/4}(\mu(x_<)^2)\mathcal{W}_{E/2\omega^2, 1/4}(\mu(x_>)^2)$$

(3.8a)

where $U(x) = \frac{1}{8m(x)}\left[m''/m - \tfrac{7}{4}(m'/m)^2\right]$.

(b) $q(x)^2 = (2\lambda/\mathcal{E})\mu(x)$:

$$V(x) = -\tfrac{\lambda}{2}\omega^2\mu(x)^{-1} - \tfrac{3}{32}\omega^2\mu(x)^{-2} + U(x)$$

$$\Leftrightarrow \bar{g}_E(x,\bar{x}) = \frac{4}{\sqrt{\pi}}\Gamma\left(\tfrac{3}{4} - \tfrac{\lambda}{\alpha_E}\right)\frac{[m(x)m(\bar{x})]^{1/4}}{\omega\alpha_E} \times$$

$$\left[\mathcal{M}_{\lambda/\alpha_E, 1/4}(\alpha_E\mu(x_<)) + \tfrac{1/4}{\sqrt{\pi}}\Gamma\left(\tfrac{1}{4} - \tfrac{\lambda}{\alpha_E}\right)\mathcal{W}_{\lambda/\alpha_E, 1/4}(\alpha_E\mu(x_<))\right]\mathcal{W}_{\lambda/\alpha_E, 1/4}(\alpha_E\mu(x_>))$$

$$\Rightarrow \bar{g}_E(x,\bar{x}) = \frac{4}{\sqrt{\pi}}\Gamma\left(\tfrac{3}{4} - \tfrac{\lambda}{\alpha_E}\right)\frac{[m(x)m(\bar{x})]^{1/4}}{\omega\alpha_E}\mathcal{M}_{\lambda/\alpha_E, 1/4}(\alpha_E\mu(x_<))\mathcal{W}_{\lambda/\alpha_E, 1/4}(\alpha_E\mu(x_>))$$

(3.8b)

where $\alpha_E = 2\sqrt{-2E}/\omega$.

Finally, we give several examples in a tabular form (Table I) for systems with different position dependent mass. For each system we write down the dimensionless integral $\mu(x)$ which is needed for the calculation of the potentials and Green's functions in (3.8a) and (3.8b). The table also lists the potential component $U(x)$ for each system. Our criterion for the selection of these mass distributions is that the square root of $m(x)$ is analytically integrable so that $\mu(x)$ is easily attainable by using Eq. (2.9). Moreover, we made an attempt to include mass functions that are frequently used in the literature. The mass distribution in the first example was studied in Ref. 7 and Ref. 11. The second example represents a smooth mass step that becomes abrupt as $\omega$ becomes large. This example was treated in Ref. 6 but for a potential that has the same shape of a smooth step. Here, exact solvability gives two systems with this mass step but for potentials that differ from the one in Ref. 6. Example three is for asymptotically vanishing mass with a maximum value of $1/\gamma$ at the origin. The mass in the fourth example is asymptotically flat with the value $m = 1$ but with a dip in the neighborhood of the origin. We show graphically, in Figs. 1 and 2, the mass and the two potential functions of the third example for several values of the dimensionless parameter $\gamma$ while keeping the other parameters fixed at $\omega = 1.0$ and $\lambda = 2.0$.

In conclusion we should point out that finding the Green's function for other potential classes is also possible. The Coulomb, Morse, Scarf, and Pöschl-Teller potentials are among such classes. The three-dimensional problem could, as well, be treated using the PCT method as outlined in Ref. 12. Furthermore, the same formalism could, in principle, be extended to include non-analytic potential classes.

**REFERENCES:**

[1] G. Bastard, *Wave Mechanics Applied to Semiconductor Heterostructure* (Les Editions de Physique, Les Ulis, France, 1988)

[2] P. Harrison, *Quantum Wells, Wires and Dots* (John Wiley and Sons, 2000); L. Serra and E. Lipparini, Europhys. Lett. **40**, 667 (1997)

[3] M. Barranco, M. Pi, S. M. Gatica, E. S. Hernandez, and J. Navarro, Phys. Rev. B **56**, 8997 (1997)

[4] F. Arias de Saavedra, J. Boronat, A. Polls, and A. Fabrocini, Phys. Rev. B **50**, 4248 (1994)

[5] See, for example, T. Gora and F. Williams, Phys. Rev. **177**, 1179 (1969); O. Von Roos, Phys. Rev. B **27**, 7547 (1983); O. Von Roos and H. Mavromatis, Phys. Rev. B **31**, 2294 (1985); R. A. Morrow, Phys. Rev. B **35**, 8074 (1987); **36**, 4836 (1987); W. Trzeciakowski, Phys. Rev. B **38**, 4322 (1988); I. Galbraith and G. Duggan, Phys. Rev. B **38**, 10057 (1988); K. Young, Phys. Rev. B **39**, 13434 (1989); G. T. Einevoll, P. C. Hemmer, and J. Thomsen, Phys. Rev. B **42**, 3485 (1990); G. T. Einvoll, Phys. Rev. B **42**, 3497 (1990); C. Weisbuch and B. Vinter, *Quantum Semiconductor Heterostructures* (Academic Press, New York, 1993)

[6] L. Dekar, L. Chetouani, and T. F. Hammann, J. Math. Phys. **39**, 2551 (1998); Phys. Rev. A **59**, 107 (1999)

[7] A. R. Plastino, A. Rigo, M. Casas, F. Gracias, and A. Plastino, Phys. Rev. A **60**, 4318 (1999); Be. Gönül, Bü. Gönül, D. Tutcu, and O. Özer, Mod. Phys. Lett. A **17**, 2057 (2002)

[8] K. Samani and F. Loran, *Shape invariant potentials for effective mass Schrödinger equation*, e-print arXiv: quant-ph/0302191

[9] V. Milanović and Z. Iković, J. Phys. A **32**, 7001 (1999)

[10] A. de Souza Dutra and C. A. S. Almeida, Phys. Lett. A **275**, 25 (2000)

[11] B. Roy and P. Roy, J. Phys. A **35**, 3961 (2002)

[12] A. D. Alhaidari, Phys. Rev. A **66**, 042116 (2002)

[13] R. Koç, M. Koca, and E. Körcük, J. Phys. A **35**, L527 (2002)

[14] L. Chetouani, L. Dekar, and T. F. Hammann, Phys. Rev. A **52**, 82 (1995)

[15] See, for example, M. F. Manning, Phys. Rev. **48**, 161 (1935); A. Bhattacharjie and E. C. G. Sudarshan, Nuovo Cimento **25**, 864 (1962); N. K. Pak and I. Sökmen, Phys. Lett. **103A**, 298 (1984); H. G. Goldstein, *Classical Mechanics* (Addison-Wesley, Reading-MA, 1986); R. Montemayer, Phys. Rev. A **36**, 1562 (1987); G. Junker, J Phys. A **23**, L881 (1990)

[16] L. Šamaj, J. Percus, and P. Kalinay, *Universal behavior of quantum Green's Functions*, e-print arXiv: math-ph/0210004

[17] H. Bateman and A. Erdélyi, *Higher Transcendental Functions* (McGraw-Hill, New York, 1953); W. Magnus, F. Oberhettinger, and R. P. Soni, *Formulas and Theorems for the Special Functions of Mathematical Physics*, 3rd edition (Springer-Verlag, New York, 1966); H. Buchholz, *The Confluent Hypergeometric Function* (Springer-Verlag, New York, 1969); I. S. Gradshtein and I. M. Ryzhik, *Table of Integrals, Series and Products* (Academic, New York, 1980)

| $m(x)$ | $\mu(x)$ | $U(x)$ |
|---|---|---|
| $\left[\dfrac{\gamma+(\omega x)^2}{1+(\omega x)^2}\right]^2$ | $\omega x+(\gamma-1)\tan^{-1}(\omega x)$ | $\omega^2\left(\dfrac{\gamma-1}{2}\right)\dfrac{3(\omega x)^4+2(2-\gamma)(\omega x)^2-\gamma}{\left[\gamma+(\omega x)^2\right]^4}$ |
| $1+\tanh(\omega x)$ | $\sqrt{2}\tanh^{-1}\left[\sqrt{1+\tanh(\omega x)}/\sqrt{2}\right]$ | $-\dfrac{\omega^2}{32}\dfrac{7+\tanh(\omega x)}{\left[\sinh(\omega x)+\cosh(\omega x)\right]^2}$ |
| $\left[\gamma+(\omega x)^2\right]^{-1}$ | $\ln\left[\omega x+\sqrt{\gamma+(\omega x)^2}\right]$ | $-\dfrac{\omega^2}{8}\dfrac{(\omega x)^2+2\gamma}{(\omega x)^2+\gamma}$ |
| $\tanh(\omega x)^2$ | $\ln\left[\cosh(\omega x)\right]$ | $-\dfrac{\omega^2}{2}\left[\sinh(\omega x)^{-2}+(5/4)\sinh(\omega x)^{-4}\right]$ |

**Table I**

**TABLE CAPTION:**

The mass function $m(x)$, the corresponding integral $\mu(x)$, and the potential component $U(x)=\dfrac{1}{8m(x)}\left[m''/m-\dfrac{7}{4}(m'/m)^2\right]$ for each the four examples mentioned at the end of Sec. III. $\omega$ is the length scale parameter and $\gamma$ is a dimensionless parameter.

**FIGURE CAPTIONS:**

FIG. 1
The mass as a function of position for the third example (third row in Table I). In the figure $\omega = 1.0$ (in arbitrary units) and the dimensionless parameter is assigned the values $\gamma = 0.2, 0.5, 1.0, 2.0, 5.0$ which are shown on their respective traces.

FIG. 2
The two potential functions for the third example with the parameter values given in the Caption of Fig. 1. The potential function given by Eq. (3.8a)/Eq. (3.8b) was used to produce Fig. 2a/Fig. 2b, respectively. We took $\lambda = 2.0$ in Fig. 2b.

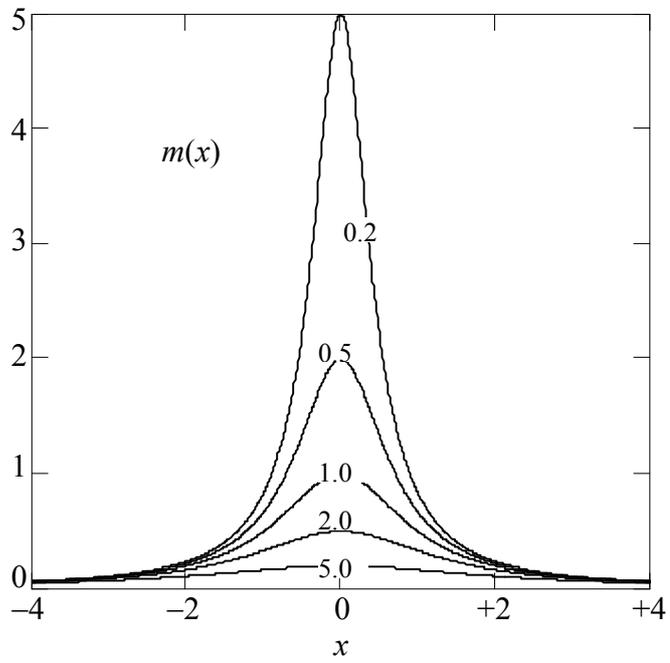

FIG. 1

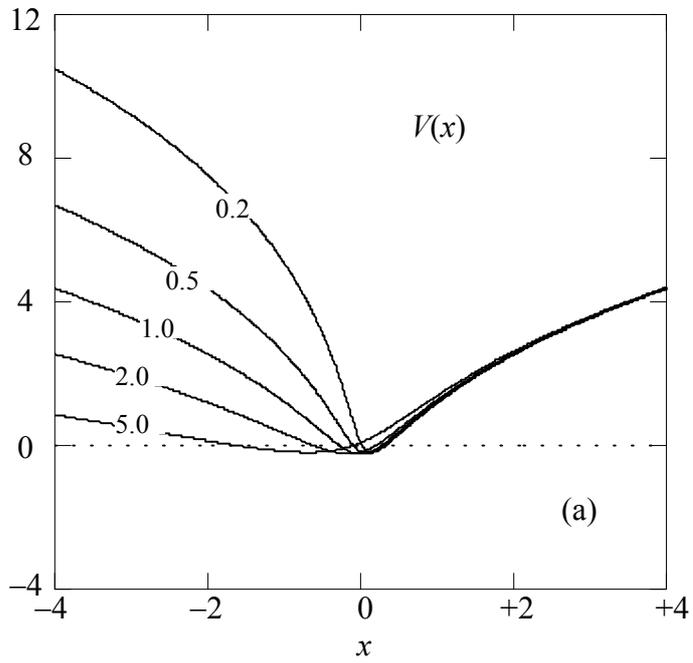

FIG. 2a

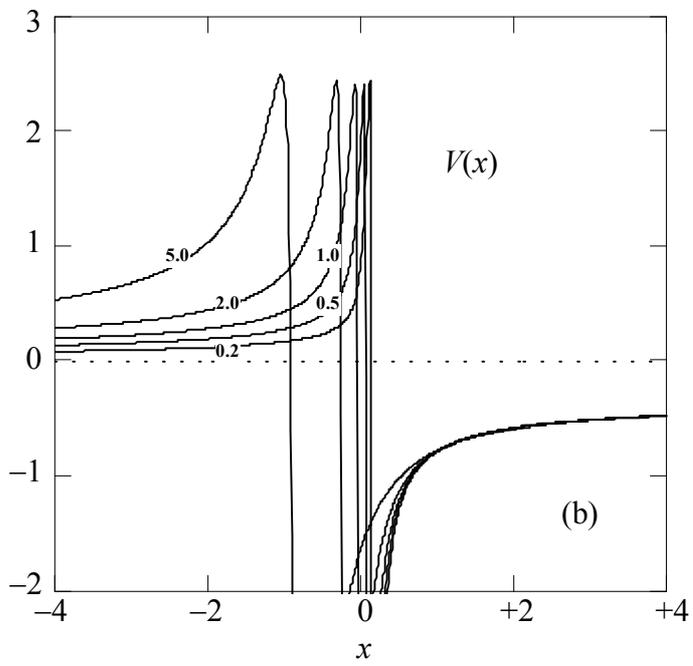

FIG. 2b